\documentclass[aps,prx,
  superscriptaddress,
  showpacs,
  twocolumn,
  nobalancelastpage,
  nofootinbib
]{revtex4-1}

\usepackage{graphicx}
\usepackage{amsmath,amssymb}
\usepackage{xcolor}
\usepackage[colorlinks=true,allcolors=blue]{hyperref}
\usepackage[normalem]{ulem}
\usepackage[bottom]{footmisc}

\newcommand{\IZ}[1]{\textcolor{orange}{#1}}

\begin{document}

\title{High-Field NMR Characterization and Indirect $J$-Spectroscopy \\ of a Nuclear Spin Chain [U-$^{13}$C,$^{15}$N]-butyronitrile}

\author{Alexey Kiryutin}
\affiliation{Novosibirsk State University, Novosibirsk, Russia}
\affiliation{International Tomography Center SB RAS, Novosibirsk, Russia}
\author{Ivan Zhukov}
\affiliation{International Tomography Center SB RAS, Novosibirsk, Russia}
\author{Danil Markelov}
\affiliation{International Tomography Center SB RAS, Novosibirsk, Russia}
\author{Erik Van Dyke}
\affiliation{Johannes Gutenberg University Mainz, Mainz, Germany}
\affiliation{Helmholtz Institute Mainz, Germany}

\author{Alexandra Yurkovskaya}
\affiliation{International Tomography Center SB RAS, Novosibirsk, Russia}

\author{Danila Barskiy}
\email{barskiy@miami.edu}
\affiliation{Department of Chemistry, University of Miami, Coral Gables, USA}
\affiliation{Frost Institute for Chemistry and Molecular Science, Coral Gables, USA}

% \authorcontributions{D.B. conceived the study. A.K., D.M., and I.Z. performed the experiments. E.V.D. contributed to data analysis and simulations. ... wrote the manuscript with input from all authors.}
% \authordeclaration{The authors declare no competing interests.}
% \equalauthors{}
% \correspondingauthor{\textsuperscript{2}To whom correspondence should be addressed. E-mail: \texttt{barskiy@miami.edu}}


\begin{abstract}
\renewcommand{\thempfootnote}{\arabic{mpfootnote}}

One-dimensional chains of coupled spins are minimal models of strongly correlated quantum matter, and have been proposed as wires for transporting quantum information. In liquids, rapid molecular tumbling averages anisotropic dipolar couplings and leaves effective isotropic scalar $J$-coupling Hamiltonians. At zero- to ultralow-field (ZULF) conditions, differences in frequency between nuclear spins of different types are quenched and the internal Hamiltonians can be closely approximated by an isotropic Heisenberg model. In this work,\footnote{The paper is based on the discussions prior to February 2022 and projects initiated at that time.} we present [U-$^{13}$C,$^{15}$N]-butyronitrile as a chemically engineered nuclear spin chain whose full spin--spin coupling network can be determined and validated by combining high-field NMR detection with evolution at ultralow fields. Starting from high-field ($16.4$ T) NMR spectra of $^1$H, $^{13}$C, and $^{15}$N nuclei, we extract all relevant $J$-couplings within a 12-spin network (four $^{13}$C, one $^{15}$N, and seven $^1$H). We then employ a mechanical field-cycling apparatus to prepolarize the spins at high field, shuttle them into a magnetically shielded region for evolution at $\lesssim$50 nT, and detect signals after returning to high field. Fourier analysis of the ultralow-field evolution yields indirect $J$-spectra that are conceptually analogous to ZULF NMR spectra but measured by a high-field NMR spectrometer. We observe clear spectral features at $J$, $1.5J$, and $2J$, in good agreement with simulations using the extracted coupling matrix. Finally, we demonstrate 2D experiments that correlate high-field chemical shifts and, thus, provide map the molecular spin chain. Our results establish [U-$^{13}$C,$^{15}$N]-butyronitrile as an extremely well-characterized spin chain model system and provide a quantitative Hamiltonian benchmark for future hyperpolarization and quantum-control studies.
\end{abstract}

%\dates{This manuscript was compiled on \today}
%\doi{\url{www.pnas.org/cgi/doi/10.1073/pnas.XXXXXXXXXX}}

\maketitle
%\thispagestyle{firststyle}

%\firstpage{1}

\section*{Introduction}

%\DB{COLOR SCHEME: 198-107-170 (purple), blue-ish (87-76-152), green-ish (76-154-141), light-green (135-161-122), orange-ish (254-199-112), pink (242-127-137)}

One-dimensional chains of coupled spins constitute one of the simplest realizations of strongly correlated quantum many-body systems \cite{christandl2004perfect,bose2007quantum,rajabi2019dynamical}. In their idealized form, such spin chains are described by Heisenberg- or XY-type Hamiltonians and have long served as canonical models to explore quantum magnetism, thermalization, and entanglement transport \cite{wang2001entanglement, schuch2003natural, wang2020xy}. Beyond their role in condensed-matter theory, spin chains are attractive resources for quantum technologies: they provide natural test beds for Hamiltonian engineering, quantum state transfer, and analog quantum simulations \cite{wei2018exploring, choi2020robust, yao2011robust, burgarth2005conclusive, bose2007quantum, georgescu2014quantum, simon2011quantum, bose2003quantum, christandl2004perfect, ajoy2013perfect}.

In the context of quantum information processing, linear spin chains have been proposed as ``quantum wires’’ that coherently relay quantum states between spatially separated registers without the need for fine-grained local control \cite{wojcik2005unmodulated, schirmer2009fast, cappellaro2007simulations}. In such architectures, an excitation initialized at one end of the chain can be transported to the opposite end by free evolution under native couplings \cite{sheberstov2025aliphatic}. These proposals have motivated a broad search for physical platforms where transport Hamiltonians arise naturally and can be controlled with a modest experimental overhead \cite{chapman2016experimental, vinet2012construct, segev2013anderson, damanet2019controlling, fujiwara2019transport, oka2019floquet}.

Solid-state NMR has provided one route to implementing and probing spin-chain physics. Dipolar-coupled nuclear spins in crystals such as fluorapatite form quasi-one-dimensional networks that approximate ideal spin-$1/2$ chains \cite{cho1996solid_MQ,cappellaro2007simulations,zhang2009nmr,ramanathan2011experimental}. Using only radiofrequency control, these systems have enabled simulations of double-quantum and XY Hamiltonians, studies of multiple-quantum coherence growth, and observation of coherent magnetization transport over tens of lattice sites \cite{cappellaro2007simulations,zhang2009nmr,ramanathan2011experimental}. However, solid-state realizations are typically constrained by disorder, dipolar inhomogeneity, and limited chemical tunability.

An appealing complementary route is offered by molecular nuclear spins in solution. In liquid-state NMR, rapid molecular tumbling averages anisotropic dipolar couplings, leaving effective Hamiltonians dominated by isotropic scalar $J$-couplings \cite{Suter2008}. At high magnetic fields, large Zeeman splittings truncate this interaction, and one usually works in the weak-coupling limit, where spins are conveniently addressable by distinct chemical shifts but the underlying Heisenberg symmetry is largely hidden. At zero- to ultralow field (ZULF) conditions, by contrast, chemical shift differences are quenched and the internal Hamiltonian of a coupled spin network reduces to 
\begin{equation} \label{Eq_1}
    \hat{H}_J = -2\pi \hbar \sum_{a<b} J_{ab}\, \hat{\mathbf{I}}_a \cdot \hat{\mathbf{I}}_b,
\end{equation}
i.e., an isotropic Heisenberg Hamiltonian on a finite graph defined by the molecular connectivity (i.e., a finite network where each nuclear spin is a node and each $J_{ab}$-coupling is an edge) \cite{ZeroFieldReview,Ledbetter2013}. In this regime, simple organic molecules with linear carbon backbones realize natural, chemically engineered nuclear spin chains whose dynamics directly reflect paradigmatic spin models \cite{sheberstov2025aliphatic}.

Zero- to ultralow-field (ZULF) NMR has recently undergone rapid development, driven by progress in optical magnetometry and alternative polarization and detection schemes \cite{ZeroFieldReview,Ledbetter2013,Eills2023HyperpolReview}. Prepolarization in a strong magnet followed by shuttling into a magnetically shielded region allows high-resolution spectroscopy of untruncated spin--spin interactions \cite{Ledbetter2013,zhukov2018field,Zhukov2021ZULF2D}. In the inverse regime, when both detection with an optically pumped magnetometer (OPM) and signal generation happen in the same location, hyperpolarization makes it possible to perform ``NMR without magnets'' \cite{Ledbetter2013,Eills2023HyperpolReview}. These advances open the door to using molecular spin networks as model systems for quantum simulation and to benchmarking quantum control protocols in the ZULF regime \cite{Jiang2018ZULFControl}.

To leverage such molecular spin chains as quantitative testbeds, one needs precise knowledge of the underlying spin Hamiltonians. In liquids, this reduces to determining the complete $J$-coupling matrix, including weak long-range interactions, and validating it against dynamics under different field conditions. High-field NMR provides exquisite spectral resolution but can obscure strongly coupled manifolds; ZULF NMR exposes the full coupling graph but often relies on non-inductive detection and specialized instrumentation \cite{ZeroFieldReview,Eills2023HyperpolReview,Zhukov2021ZULF2D}. A combined approach that uses conventional high-field spectra to extract couplings, and ultralow-field evolution to cross-check and refine them, is therefore attractive.

%\DB{In this work, we realize such an approach using [U-$^{13}$C,$^{15}$N]-butyronitrile, a simple organic molecule with a four-spin $^{13}$C backbone terminated by a $^{15}$N and coupled to seven protons. This system forms a finite nuclear spin chain that is chemically relevant yet sufficiently small to permit exact spin-system simulations. We first use high-field ($16.4$ T) $^1$H, $^{13}$C, and $^{15}$N spectra, fitted with ANATOLIA software \cite{ANATOLIA}, to extract all relevant $J$-couplings with $\sim$0.1~Hz accuracy. We then employ a mechanical field-cycling apparatus spanning 50 nT to 9.4\,T \cite{zhukov2018field} mounted on a 400 MHz spectrometer, to perform prepolarization at high field, ultralow-field evolution at $\lesssim 50$ nT, and high-field detection on the same sample. Fourier analysis of the ultralow-field kinetics leads to indirect $J$-spectra of the spin chain that are conceptually similar to ZULF NMR spectra but measured via high-field inductive detection. Finally, we demonstrate 2D correlation experiments, spectroscopic maps of quantum coherence through the entire molecular spin chain.}

In this work, we present a complete reconstruction of a 12-spin liquid-state molecular spin-chain Hamiltonian using uniformly labeled [U-$^{13}$C,$^{15}$N]-butyronitrile as a model system. By combining multinuclear high-field NMR spectroscopy, used to determine all scalar couplings with high precision ($<$0.05 Hz), with controlled evolution at zero to ultralow magnetic fields enabled by magnetic field cycling, we establish and validate the effective Heisenberg character of the spin chain. Fourier analysis of ultralow-field dynamics yields indirect $J$-spectra that are conceptually equivalent to ZULF NMR $J$-spectra, while relying solely on standard high-field inductive detection, thereby bridging conventional high-field and emerging zero-field modalities within a single experimental platform. This approach constitutes a broadly applicable Hamiltonian tomography protocol suitable for analysis of other solution-state molecular spin chains and it provides a rigorous foundation for future hyperpolarization-enhanced NMR experiments (e.g., via SABRE) aimed at initializing and exploiting non-equilibrium spin states for molecular quantum-simulation applications.

%This thermal-polarization study fully characterizes the [U-$^{13}$C,$^{15}$N]-butyronitrile spin-chain Hamiltonian and bridges the gap between the conventional high field and emerging zero- to ultra low field NMR modalities via employing field-cycling methodology. The resulting Hamiltonian forms the foundation for hyperpolarized experiments on the same molecule, where parahydrogen-based methods such as SABRE can be used to initialize highly ordered, non-equilibrium spin states for quantum-simulation protocols. Those experiments will be reported separately.

%\AK{In this work, we present the first complete reconstruction of a 12-spin liquid-state molecular spin-chain Hamiltonian and validate its Heisenberg character via evolution at zero to ultralow magnetic fields. By combining high-field multi-nuclear spectra with ultralow-field dynamics, we demonstrate a broadly applicable Hamiltonian tomography protocol for chemically engineered spin chains in solution. Conceptually, our approach introduces indirect J-spectroscopy — effectively ‘ZULF NMR on a high-field spectrometer’ — which leverages standard hardware to access zero-field-like information content.}

%\IZ
%{ % IZ 24/12/2025

    \section*{Methods}
    \subsection*{Sample preparation and NMR experiments}

    [U-$^{13}$C, $^{15}$N]-butyronitrile was obtained from Sigma-Aldrich via custom synthesis.
    For the NMR experiments, we used solutions of [U-$^{13}$C, $^{15}$N]-butyronitrile at 230 mM in methanol-$d_4$. The solution was subjected to three freeze--pump--thaw cycles to remove dissolved oxygen and sealed in standard 5 mm NMR tubes. High-resolution spectra for spin-system simulations were acquired on a spectrometer with a 700 MHz base proton frequency at ambient temperature (298~K). % Is it a correct temperature?
    
    Ultralow-field experiments were performed on a home-built mechanical field-cycling setup based on a 400 MHz NMR spectrometer with a magnetic shield installed above the cryomagnet \cite{zhukov2018field}. The protocol consisted of prepolarization at 9.4 T, adiabatic shuttling to 50 $\mu$T, a rapid field switch to $B \lesssim$50~nT for evolution during a variable time $\tau$, followed by adiabatic return to 9.4 T and FID detection of $^1$H, $^{13}$C, or $^{15}$N NMR signals. % \DB{(total number of points sampled at ZULF conditions was ...)}
    
    \subsection*{$J$-coupling determination from high-field spectra}
    
    High-resolution $^1$H, $^{13}$C, and $^{15}$N NMR spectra at 16.4 T were recorded using standard 1D acquisition with appropriate decoupling schemes. Spectra were fitted using the ANATOLIA software package \cite{ANATOLIA}, which performs least-squares optimization of chemical shifts and $J$-couplings based on density-matrix simulations of the full spin system. Protons in the CH$_3$ and CH$_2$ groups were treated as magnetically equivalent (A$_3$M$_2$X$_2$ model), and intra-group couplings were neglected as they do not affect the spectra at the level of precision considered. The error of the $J$-couplings determined in this way is estimated to be $\sim 0.01$--0.1 Hz.
    
    \subsection*{Data analysis and simulations}
    
    Time-domain data $S(\tau)$ acquired at ultralow field were apodized with exponential or Gaussian windows, zero-filled, and Fourier transformed to obtain indirect $J$-spectra. 2D ZULF-TOCSY correlation spectra were processed in Topspin 3.7 and/or Mnova 12.0 \cite{mnova} using conventional 2D NMR routines.  Spin-system simulations were performed using custom scripts \cite{iRelax} implementing the full 12-spin Hamiltonian, with $J$-couplings fixed to the values obtained from high-field fits. Relaxation effects were excluded from calculations in iRelax \cite{iRelax}.

%} % IZ 24/12/2025

\section*{Results and Discussion}

\subsection*{Molecular spin chain and high-field NMR spectra}

\begin{figure}[t]
    \includegraphics[width=1\columnwidth]{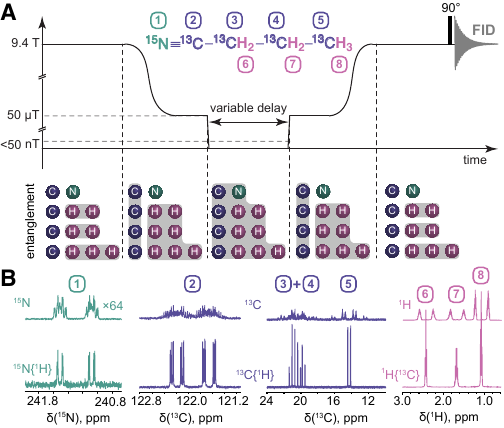}
    \caption{(A) Molecular structure of [U-$^{13}$C,$^{15}$N]-butyronitrile and a magnetic field cycling protocol used in this work. The protocol consists of the following stages: prepolarization at 9.4 T, adiabatic shuttling to 50 $\mu$T, sudden (100 $\mu$s) switch to ultralow field ($<$50 nT) for a variable time $\tau$, and subsequent adiabatic return to the high field for the free induction decay (FID) detection. At each stage, different nuclear spins are entangled, as shown by the gray-shaded areas: e.g., only equivalent $^{1}$H nuclei are entangled at 9.4 T, whereas all nuclei are entangled at 50~nT. (B) Selected regions of high-field (9.4~T) $^1$H, $^{13}$C, and $^{15}$N NMR spectra with and without decoupling.} 
    %(D) Representative kinetics of NMR signals from different nuclei as a function of evolution time at ultralow field.}
    \label{fig1}
\end{figure}

Figure~\ref{fig1}A shows the structure of a molecular spin chain [U-$^{13}$C,$^{15}$N]-butyronitrile and the labeling of distinct nuclear spin types. The four $^{13}$C spins form a linear backbone (C$_2$--C$_5$) terminated by a $^{15}$N (N$_1$), while protons (H$_6$--H$_8$) represent an aliphatic spin chain CH$_2$--CH$_2$--CH$_3$. The protons within each CH$_3$ and CH$_2$ group are chemically equivalent and the molecular motif in question can be treated as a A$_3$M$_2$X$_2$ sub-system in the spin-system fits.

\begin{figure*}[ht]
\centering
\includegraphics[width=2\columnwidth]{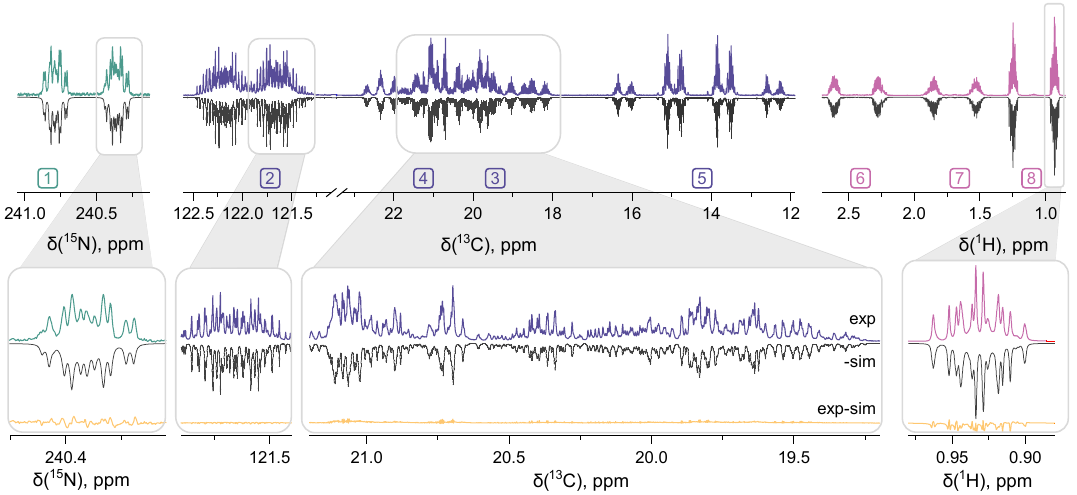} 
\caption{Experimental (colored) and simulated (black) $^1$H, $^{13}$C, and $^{15}$N NMR spectra of [U-$^{13}$C,$^{15}$N]-butyronitrile at 16.4 T fitted with ANATOLIA \cite{ANATOLIA}. Insets show residuals (experiment minus simulation).}
\label{fig2}
\end{figure*}

High-resolution $^1$H, $^{13}$C, and $^{15}$N NMR spectra were recorded at 16.4 T ($^1$H Larmor frequency 700 MHz) using a standard BBO probe (Fig.~\ref{fig2}). The spectra display well-resolved multiplet patterns reflecting the network of one-bond and long-range $J$-couplings. Using the ANATOLIA software package \cite{ANATOLIA}, we simultaneously fitted the $^1$H, $^{13}$C, and $^{15}$N NMR spectra, adjusting the common set of $J$-couplings until excellent agreement between experiment and simulation was obtained (Fig.~\ref{fig2}, orange residuals). The resulting coupling constants are listed in Table~\ref{tab:Jmatrix}. The uncertainties estimated from the fits are on the order of 0.1 Hz.

The extracted $J$-coupling matrix confirms the expected spin-chain topology: large one-bond $^{13}$C--$^{13}$C couplings (33--55~Hz) along the backbone, strong one-bond $^{13}$C--$^1$H couplings (130--135 Hz), and sizeable one-bond $^{15}$N--$^{13}$C ($\sim$17~Hz) and three-bond $^{15}$N--$^1$H ($\sim$2~Hz) couplings at the nitrile end. Weaker two- and three-bond $^{13}$C--$^1$H couplings (typically a few Hz) are also resolved; they play an important role in the detailed structure of the zero-field $J$-spectrum. Therefore, fully $^{13}$C- and $^{15}$N-labeled butyronitrile with all $J$-couplings resolved is now a well-defined 12-spin network with a precisely known nuclear spin Hamiltonian.

\begin{table}[]
\centering
\caption{\label{tab:Jmatrix}$J$-couplings between the spins (see labeling in Fig.~\ref{fig1}A) of [U-$^{13}$C,$^{15}$N]-butyronitrile extracted from high-field NMR spectra. Uncertainties are estimated as $\sim$0.05 Hz.}
\begin{tabular}{c|r|r|r|r|r|r|r}
$J$ (Hz) & C$_2$ & C$_3$ & C$_4$ & C$_5$ & H$_6$ & H$_7$ & H$_8$ \\ \hline
N$_1$  & 17.37 & -2.89 & 0.58 & 0.00 & 1.65 & 0.03 & 0.00 \\
C$_2$  &   & 55.15 & -2.84 & 3.47 & -9.57 & 6.35 & 0.00 \\
C$_3$  &   &   & 32.97 & -1.02 & 135.2 & -4.15 & 6.4 \\
C$_4$  &   &   &   & 34.81 & -5.07 & 130.61 & -4.48 \\
C$_5$  &   &   &   &   & 4.83 & -4.24 & 126.09 \\
H$_6$  &   &   &   &   &   & 7.02 & -0.04 \\
H$_7$  &   &   &   &   &   &   & 7.41 \\ 
\end{tabular}
\end{table}

\subsection*{Field-cycling apparatus and ultralow-field evolution}

To probe the dynamics of this spin chain under an effectively isotropic Heisenberg Hamiltonian, we employ a mechanical field-cycling setup built on a 400 MHz NMR spectrometer \cite{zhukov2018field}. As summarized in Fig.~\ref{fig1}A, the apparatus allows positioning of the sample anywhere between the homogeneous bore of the superconducting magnet (up to 9.4 T) and a magnetically shielded region above the magnet where the field can be controlled in the 50~nT--2~mT range using a coil system \cite{zhukov2018field}. Key features of the setup include a continuous field range spanning $\sim$9 orders of magnitude (50 nT--9.4 T), field-switching times between high field and the shielded region below 0.5~s, non-adiabatic field jump inside the magnetic shield in the 50~nT--2~mT region within $\lesssim$100~$\mu$s, high effective spectral resolution ($\sim$0.5~Hz, corresponding to $\sim$1~ppb at 9.4~T) in the high-field detection stage, compatibility with multinuclear probes (BBO, TXI) enabling detection of $^1$H, $^{13}$C, $^{15}$N, and other nuclei, robust mechanical performance, with $\sim$$10^6$ shuttling cycles over several years of operation.
    
For the thermally polarized NMR experiments reported here, samples of [U-$^{13}$C,$^{15}$N]-butyronitrile (230~mM in methanol-$d_4$) were prepolarized at 9.4~T for, %\DB{several $T_1$ periods (what exact time, T1 of which nuclei?)}
then shuttled adiabatically to a field of 50 $\mu$T in the shielded region. After stabilization, a sudden (100 $\mu$s) field switch reduced the field to $B \lesssim 50$ nT, at which point the spin system evolved under the isotropic $J$-Hamiltonian for a variable evolution time $\tau$. The sample was then shuttled back to the high-field position for free induction decay (FID) detection of $^1$H, $^{13}$C, or $^{15}$N signals (Fig.~\ref{fig1}B).

Representative kinetics of the signal intensities as a function of $\tau$ are shown in Fig.~\ref{fig3} (left panels). The signals exhibit coherent oscillations at frequencies determined by the eigenvalue differences of the zero-field $J$-Hamiltonian, damped by relaxation and by residual inhomogeneities in the ultralow field. The oscillation frequencies differ for each observed nucleus, reflecting their distinct coupling environments within the spin chain.

\subsection*{Indirectly measured $J$-spectra}

\begin{figure}[b]
\centering
\includegraphics[width=\columnwidth]{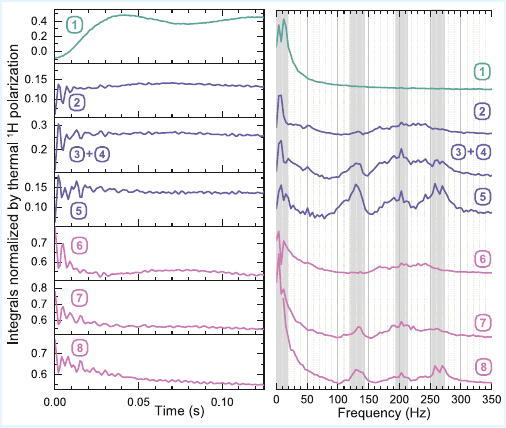}
\caption{(Left) Evolution of NMR signal intensities from different nuclei as a function of evolution time at 50 nT. (Right) Indirectly measured $J$-spectra obtained by Fourier transformation of the signals on the left. Vertical gray boxes indicate approximate positions of $J$-, $1.5J$-, and 2$J$-peaks, where $J$ is the most relevant heteronuclear single-bond $J$-coupling. As in conventional ZULF NMR spectra, features at $J$ and $2J$ are observed \IZ{f}or $^{13}$CH$_3$ group (subplots 8 and 5), and $3J/2$ peak is observed for $^{13}$CH$_2$ groups (subplots 3+4, 6 and 7).}
\label{fig3}
\end{figure}

To obtain indirect $J$-spectra, we compute the Fourier transform of the time traces $S(\tau)$ collected for each nucleus at a fixed ultralow field (here 50 nT), after appropriate windowing and zero-filling. The resulting spectra are shown in Fig.~\ref{fig3} (right). They display peaks at frequencies corresponding to allowed transitions between eigenstates of $\hat{H}_J$, and are conceptually analogous to ZULF NMR spectra recorded with non-inductive detection \cite{ZeroFieldReview,Ledbetter2013}, but here they are measured by a high-field NMR spectrometer equipped with fast field-cycling setup.

For the $^{13}$CH$_3$ group, prominent peaks appear at $J_{\mathrm{CH}}$ and $2J_{\mathrm{CH}}$, highlighting the dominant one-bond $^{13}$C--$^1$H coupling. For the $^{13}$CH$_2$ group, the strongest feature is at $\tfrac{3}{2}J_{\mathrm{CH}}$, as expected from the eigenvalue structure of the XA$_2$ sub-system \cite{ZeroFieldReview}. Additional weaker lines arise from longer-range couplings and from interactions with the $^{15}$N. The observed frequencies and relative intensities are in excellent agreement with simulations based on the $J$-coupling matrix extracted from high-field spectra (Table~\ref{tab:Jmatrix}), confirming the internal consistency of the proposed Hamiltonian structure (Figs. S1-S2).

The indirect $J$-spectra obtained in this way have several noteworthy features. First, they are measured using commercially available high-resolution high-field NMR hardware with inductive (Faraday) detection, without the need for atomic magnetometers and/or SQUIDs \cite{Ledbetter2013}. Second, the spectral resolution in the frequency domain is set by the maximum evolution time at ultralow field, which was 0.128 s here to shorten the experimental time, but in optimal scenarios can reach several seconds before relaxation dominates, readily enabling sub-hertz resolution. Third, the method is naturally multinuclear: by picking up different nuclei at high field, one selectively probes different local environments within the spin chain. Commercially available multiple receiver NMR machines allow for the simultaneous detection of different nuclei, thus, significantly shortening the required instrument time \cite{Kupvce2021parallel,Anisimov2026simultaneous}.

\subsection*{Zero-/high-field 2D NMR total correlation spectra}

\begin{figure*}[t]
    \includegraphics[width=2\columnwidth]{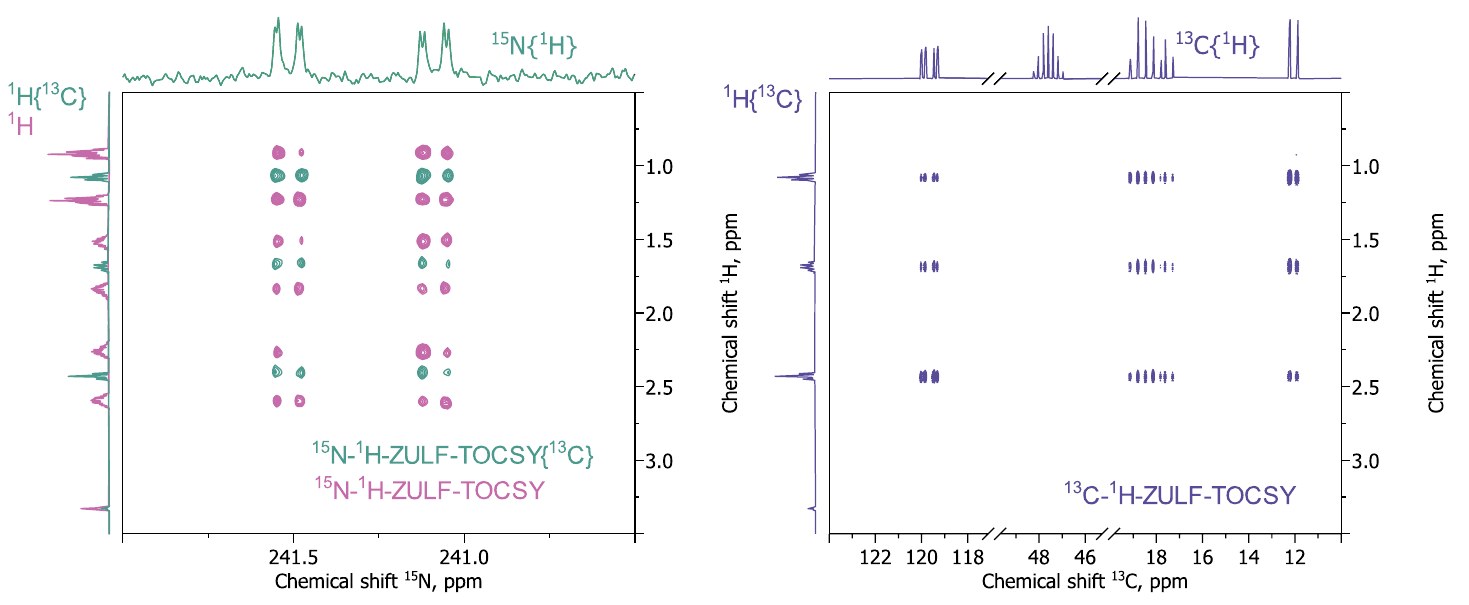}
    %\caption{(left) $^{13}$C-$^{1}$H-ZULF-TOCSY NMR spectra, (right) $^{15}$N-$^{1}$H-ZULF-TOCSY NMR spectrum. make caption for two columns}
    \caption{(Left) $^{15}$N-$^{1}$H-ZULF-TOCSY NMR spectra, taken with (red) and without (green) 180-degree refocusing pulse at $^{13}$C radio-frequency channel in the middle of evolution period $t_1$. Experimental parameters: relaxation delay - 6~s, transfer time (one direction) - 479~ms, mixing time at ZULF conditions $t_{\rm mix} =$ 50~ms, ZULF evolution field $\lesssim$-50 nT, TD($^{1}$H) = 128, TD($^{15}$N) = 2048, States-TPPI acquisition mode. Processing parameters: pure cosine-squared window function in both dimensions, SI($^{1}$H) = 512, SI($^{15}$N) = 4096. (Right) $^{13}$C-$^{1}$H-ZULF-TOCSY NMR spectrum. Experimental parameters: relaxation delay - 6 s, transfer time (one direction) - 479 ms, mixing time at ZULF conditions $t_{\rm mix} =$ 10~ms, ZULF evolution field - $\lesssim$50 nT, TD($^{1}$H) = 256, TD($^{13}$C) = 32768, States-TPPI acquisition mode. Processing parameters: pure cosine-squared window function in both dimensions, SI($^{1}$H) = 512, SI($^{13}$C) = 4096.}
    \label{fig4}
\end{figure*}

To test the performance of [U-$^{13}$C,$^{15}$N]-butyronitrile strongly coupled spin chain as a wire for quantum spin order transport, we performed 2D NMR experiments in which the system evolved for a variable time $t_1$ under the high-field Hamiltonian, transferred to the magnetic shield for evolution under the zero- to ultralow-field Hamiltonian for a fixed mixing time $t_{\rm mix} =$ 10-50~ms, transferred back, and evolved for a time $t_2$ under high-field conditions during detection (see Fig.~\ref{fig4}). The resulting 2D spectra $S(\omega_{1}^\mathrm{HF},\omega_{2}^\mathrm{HF})$, correlate high-field resonance frequencies of all protons 6, 7, and 8 in indirect dimension with all high-field resonance frequencies of the directly observed heteronuclei ($^{13}$C or $^{15}$N) \cite{zhukov2020total} indicating the effective transfer of the information encoded in proton magnetization via the strong one-bond $J$-couplings network over the entire spin chain. It is important to stress out that additional mixing of spin states within each nuclear spin subspace---protons 6-7-8 and carbons 2-3-4-5---proceeds naturally during the sample transfer time ($\sim$0.5~s) between the magnetic shield and the detection position of the high field NMR spectrometer (Fig. 1A). The reason for mixing delay $t_{\rm mix}$ at the zero- to ultra low field is to entangle different nuclear spin subspaces, in this case, all the magnetic nuclei: $^{1}$H, $^{13}$C, and $^{15}$N (see kinetics of the signal intensities shown in Fig.~\ref{fig3}).

%\DB{Discuss Fig. S1 in more detail, cross-relaxation vs. possibility of coherent transport; what is the main reason of discrepancy between expt. and simulation?}

Imposing ZULF conditions is not the only way to implement evolution under the Heisenberg Hamiltonian. Fulfilling the cross-polarization condition 
% \begin{equation2}
%     \omega_S = \gamma_{S} \cdot B_{1,S} = \omega_I = \gamma_I \cdot B_{1,I}
% \end{equation2} 
for the appropriate amplitudes 
% \begin{equation3}
%     B_{1,S}
% \end{equation3}
% and 
% \begin{equation4}
%     B_{1,I}
% \end{equation4}
of spin-locking transverse resonant magnetic fields for two (or more) heteronuclei results in the same form of effective spin Hamiltonian as Eq.~\eqref{Eq_1}, as was shown by Hartmann and Hahn \cite{Hartmann1962double}. Since then, more elaborate heteronuclear cross-polarization pulse sequences were developed for special cases, for example, solid-state NMR of biomolecules \cite{Jain2012crosspol} and liquid crystals \cite{Dvinskikh2006heteronuclear}. Another way of mixing nuclear spin subspaces by utilizing the quasi-resonant adiabatic frequency-swept pulses in the low-field range of about a few mT was demonstrated recently by Zhang and Hilty \cite{Zhang2025adiabatic}.

In the case of [U-$^{13}$C,$^{15}$N]-butyronitrile, the 2D spectra exhibit well-resolved ridges corresponding to specific groups of transitions, which can be assigned to particular segments of the spin chain. For example, transitions dominated by the C$_3$--H$_6$ coupling correlate with the C$_3$ chemical shift region, whereas those involving the nitrile end correlate with N$_1$ and C$_2$ resonances (see Fig.~\ref{fig4}). The 2D maps thus provide a powerful way to visualize propagation of polarization within the molecule. With the current field-cycling setup design, the shortest transfer time between the high-field center and the magnetic shield is about $\sim$0.5~s. If the transfer time becomes $\lesssim$0.1~s, it would be competitive with propagation of nuclear spin polarization through the coupled proton spin subsystem. To observe the same effect for carbon subsystem, transfer time less than 0.01~s would be required. In this case, 2D ZULF-TOCSY NMR spectra recorded with different transfer times would measure ultimate  speed limits imposed on nuclear spin polarization transfer.

\subsection*{Comparison with simulations and sensitivity to weak couplings}

Using the $J$-coupling matrix from Table~\ref{tab:Jmatrix}, we performed full quantum simulations of the 12-spin system under the experimental pulse sequence, including prepolarization, sudden transfer, and evolution at ultralow field. The simulated kinetics and indirect $J$-spectra reproduce the main experimental features with high fidelity, including the positions and relative amplitudes of the lines in Fig.~\ref{fig3}. Small discrepancies can be traced to uncertainties in very weak couplings ($|J|\lesssim 0.1$ Hz) and to relaxation processes not included in the model.

%\IZ{\sout{Importantly, the $J$-spectra are sensitive to these weak couplings: varying a given long-range $J$ by $\pm 0.2$ Hz in the simulations produces noticeable shifts or splittings in specific peaks. In principle, this sensitivity suggests that the combination of high-field fitting and ultralow-field evolution can be used as a robust Hamiltonian tomography protocol for molecular spin chains. One can envision the potential of such methods to refine even mHz-level couplings.}}

%\section*{Discussion} %\IZ
\section*{Conclusions} %\IZ

Our results establish [U-$^{13}$C,$^{15}$N]-butyronitrile as a quantitatively characterized nuclear spin chain and demonstrate that standard high-field NMR hardware, augmented by a robust mechanical field-cycling setup, can access the rich zero- to ultralow-field $J$-spectroscopy of such systems. The indirect $J$-spectra and 2D ZULF-TOCSY experiments provide complementary information to conventional high-field spectra and can be interpreted using a single underlying Hamiltonian.

From the perspective of quantum simulation, the system studied here represents a chemically relevant realization of a finite Heisenberg spin chain at room temperature. The fully determined $J$-matrix enables direct comparison between experiment and theoretical models of transport, decoherence, and entanglement growth in such chains. While the present work relies on thermal polarization, the same molecular platform and field-cycling apparatus are naturally compatible with hyperpolarization methods, including parahydrogen-based techniques such as SABRE \cite{Hovener2018PHIPBiomedicine, barskiy2019sabre, Eills2023HyperpolReview}. In that regime, one can initialize the spin chain in highly non-equilibrium multi-spin states and explore quantum-control protocols that go beyond the high-temperature limit. Thus, the Hamiltonian benchmark established here provides a quantitative foundation for the forthcoming hyperpolarization studies.
In the long term, bio-synthetic routes to isotopically labeled spin chains (e.g., enzyme-driven $^{13}$C and/or $^{15}$N incorporation via metabolic pathways) could dramatically reduce the cost and broaden the chemical diversity of molecular nuclear-spin-based quantum simulation platforms.

\subsection*{Acknowledgment}
D.B. thanks Prof. Dmitry Budker and Prof. Ferdinand Schmidt-Kaler for stimulating discussions. A.K., D.M., I.Z., and A.Y. acknowledge the Ministry of Science and Higher Education of the Russian Federation for granting access to the NMR equipment.

\bibliography{Bibliography}

\end{document}